\documentclass[12pt,preprint]{aastex}
\usepackage{natbib}
\usepackage{graphicx}
\shorttitle{Two analytic relations for galaxy clusters}
\begin{document}
\title{Two analytic relations connecting the hot gas astrophysics with the cold dark matter model for galaxy clusters}
\author{Man Ho Chan}
\affil{Department of Science and Environmental Studies, The Education University of Hong Kong, Hong Kong, China}
\email{chanmh@eduhk.hk}

\begin{abstract}
Galaxy clusters are good targets for examining our understanding of cosmology. Apart from numerical simulations and gravitational lensing, X-ray observation is the most common and conventional way to analyze the gravitational structures of galaxy clusters. Therefore, it is valuable to have simple analytical relations that can connect the observed distribution of the hot, X-ray emitting gas to the structure of the dark matter in the clusters as derived from simulations. In this article, we apply a simple framework that can analytically connect the hot gas empirical parameters with the standard parameters in the cosmological cold dark matter model. We have theoretically derived two important analytic relations, $r_s \approx \sqrt{3}r_c$ and $\rho_s \approx 9\beta kT/8 \pi Gm_gr_c^2$, which can easily relate the dark matter properties in galaxy clusters with the hot gas properties. This can give a consistent picture describing gravitational astrophysics for galaxy clusters by the hot gas and cold dark matter models.        
\end{abstract}

\keywords{dark matter}

\section{Introduction}
Analyzing galaxy clusters is an important way to study the gravitational physics and structure formation in cosmology. The gravitational interplay between dark matter and baryons plays a crucial role in describing the structures of galaxy clusters. Generally speaking, most of the studies for galaxy clusters rely on X-ray observations \citep{Reiprich,Balogh,Chan,Chan2}, numerical simulations \citep{Munari,Henden,Qiu} and gravitational lensing \citep{Fong,Jauzac,Meneghetti}. In particular, X-ray observation of the hot gas is the most conventional way to study galaxy clusters, which can tell us about the hot gas distribution and probe the dark matter content.

Current X-ray observations can accurately measure the brightness profiles of hot gas in many galaxy clusters \citep{Reiprich2,Chen,Cavagnolo}. The hot gas distribution can be approximately characterized by a few parameters, such as the average temperature $T$, the core radius $r_c$, the central electron number density $n_0$ and the hot gas density index parameter $\beta$ \citep{Reiprich2,Chen}. The standard framework of the hot gas analysis (the hot gas framework) can probe the dark matter content based on these parameters \citep{Reiprich2,Chen}. On the other hand, the standard cold dark matter (CDM) model can also predict the functional form of the dark matter density profile in galaxy clusters. Numerical simulations suggest that the dark matter density profile can be described by the universal Navarro-Frenk-White (NFW) profile \citep{Navarro}:
\begin{equation}
\rho_{DM}(r)=\frac{\rho_s r_s^3}{r(r+r_s)^2},
\end{equation}
where $\rho_s$ and $r_s$ are the scale density and scale radius of the dark matter density respectively. However, it seems that there is no trivial relation connecting the parameters obtained from the hot gas framework with that obtained from the CDM model (e.g. the relation between $r_c$ and $r_s$). Only some rough relations involving a few uncertain variables (e.g. the polytropic index) have been developed to connect these parameters together \citep{Suto,Komatsu,Ascasibar}.

In this article, we analytically derive two important relations that can connect the parameters in the hot gas framework with that in the CDM model. Although we have applied a few assumptions which can make the relations simple and more general, all these assumptions are indeed good approximations in general. The resultant relations can provide an easy way to connect the empirical data from X-ray observations to the theories in the CDM model. 

\section{The analytic framework}
We start with the NFW profile obtained from the CDM model. We assume that the baryonic mass components roughly follow the dark matter distribution. Therefore, by Eq.~(1), the total enclosed mass profile is given by
\begin{equation}
M(r) \approx \int_0^r4 \pi r'^2 \rho_{DM}(r')dr'=4\pi \rho_sr_s^3 \left[\ln \left(1+\frac{r}{r_s} \right)-\frac{r}{r+r_s} \right].
\end{equation}
If the hot gas is in hydrostatic equilibrium, we have 
\begin{equation}
\frac{d}{dr}[n(r)kT]=-\frac{GM(r)m_gn(r)}{r^2},
\end{equation}
where $n(r)$ is the hot gas number density profile, $m_g$ is the average mass of a hot gas particle and $T$ is the hot gas temperature. We can write Eq.~(3) in terms of the number density gradient and the temperature gradient: 
\begin{equation}
\frac{d \ln n(r)}{d \ln r}+ \frac{d \ln T}{d \ln r}=-\frac{GM(r)m_g}{kTr}.
\end{equation}
In general, the hot gas temperature is a function of $r$. The hot gas temperature has relatively large variation in the central regions of the cool-core clusters \citep{Vikhlinin,Chen,Hudson,Reiprich} and the outer regions of most galaxy clusters \citep{Vikhlinin,Zhang1,Pratt,Zhang2}. For simplicity, we assume that $T$ is constant and the value of $T$ can be represented by an average hot gas halo temperature for each galaxy cluster \citep{Chen}. Observations show that the central temperature variations are less than 10\% and 25\% for the hot gas in non-cool-core cluster and cool-core clusters respectively \citep{Hudson}. In the followings, we will examine the region near $r=r_s$ in which the temperature variation is not large. Also, the size of the cool core in a galaxy cluster is usually smaller than $r_s$. Therefore, the effect of the temperature variation in the cool-core regions is not important to our analysis. For the temperature decline in the outer regions of most galaxy clusters, we will estimate the systematic uncertainty of the assumption of constant temperature later. By combining Eq.~(2) and Eq.~(4) with $d \ln T/d\ln r=0$, we can get an analytic solution for $n(r)$:
\begin{equation}
n(r)=n_0 \exp \left[ -\frac{4\pi \rho_sr_s^2Gm_g}{kT} \left(1-\frac{\ln(1+x)}{x} \right) \right],
\end{equation}
where $x=r/r_s$. If we write $C=4\pi \rho_sr_s^2Gm_g/kT$, the above expression can be written as
\begin{equation}
n(r)=n_0 \left[\exp \left(1-\frac{\ln(1+x)}{x} \right) \right]^{-C}.
\end{equation}
Note that Eq.~(6) is the theoretical prediction of the hot gas density profile based on the CDM model, which has been derived already by some previous studies \citep{Suto}.

On the other hand, from X-ray observations, most of the hot gas density profile in galaxy clusters can be approximately expressed by the $\beta$-model \citep{Reiprich2,Chen}:
\begin{equation}
n(r)=n_0 \left(1+\frac{r^2}{r_c^2} \right)^{-3\beta/2}.
\end{equation}
Following the same assumption of hydrostatic equilibrium, we can get the enclosed dark matter mass profile based on the hot gas framework by substituting Eq.~(7) into Eq.~(4):
\begin{equation}
M(r)=\frac{-kTr}{Gm_g} \left[\frac{-3\beta r^2}{(r^2+r_c^2)}+\frac{d\ln T}{d\ln r}\right].
\end{equation}
For the sake of completeness, we keep the temperature gradient term in Eq.~(8). Hence, the dark matter density profile is
\begin{equation}
\rho_{DM}(r)=\frac{1}{4\pi r^2} \frac{dM(r)}{dr}=\frac{kT}{4\pi Gm_g} \left[\frac{3\beta(3r_c^2+r^2)}{(r_c^2+r^2)^2}+\frac{\gamma}{r^2} \left(\frac{3\beta r^2}{r^2+r_c^2}-1-\gamma \right)-\frac{1}{r}\frac{d \gamma}{dr} \right],
\end{equation}
where $\gamma \equiv d \ln T/d\ln r$. Note that the above dark matter density profile in Eq.~(9) is based on the hot gas framework while the one in Eq.~(1) is based on the CDM model. Also, as mentioned above, the variation of temperature may become large in the outer region ($r>0.1r_{200}$, where $r_{200}$ is the virial radius of a galaxy cluster) \citep{Pratt}. 

We first consider the simplest case $\gamma=0$ (i.e. constant temperature profile). In Fig.~1, we plot the dark matter density profiles based on Eqs.~(1) and (9) for comparison. We can see that deviations mainly occur at $x<0.5$ and $x>3$. If we assume that Eq.~(1) and Eq.~(9) are approximately equal (or very close) to each other at $r=r_s$, we get
\begin{equation}
\rho_s=\frac{3 \beta kT}{\pi Gm_gr_s^2}\left[ \frac{a^2(a^2+3)}{(a^2+1)^2} \right],
\end{equation}
where $a=r_s/r_c$. Interestingly, from Eq.~(10), we get
\begin{equation}
C=\frac{4 \pi \rho_sr_s^2Gm_g}{kT}=\frac{12 \beta a^2(a^2+3)}{(a^2+1)^2}.
\end{equation}
Also, from Eq.~(7) (the hot gas framework), we can see $n(r) \propto r^{-3\beta}$ for $r \gg r_c$. However, for a physical range of the hot gas size $r=2r_c-10r_c$, the power-law index is slightly smaller than 3: $n(r) \propto r^{-2.8 \beta}$. On the other hand, if we start with Eq.~(6) (the CDM framework), the function $\exp[1-\ln(1+x)/x] \propto x^{0.208}$ for $x \ge 1$. Therefore, if both frameworks are consistent with each other, we get
\begin{equation}
2.8 \beta=0.208C=0.208 \left[ \frac{12 \beta a^2(a^2+3)}{(a^2+1)^2} \right].
\end{equation}
Surprisingly, since $0.208(12)/2.8 \approx 8/9$, Eq.~(12) can be simplified to $(a^2-3)^2=0$, which gives the double root $a^2=3$. Therefore, we have 
\begin{equation}
r_s \approx \sqrt{3}r_c. 
\end{equation}
This relation is universal for galaxy clusters as it does not depend on any other hot gas parameters or dark matter parameters. In Fig.~2, we also show the hot gas number density profiles for both the hot gas framework and the CDM framework for comparison. As the range of $\beta$ is not large ($\beta=0.65 \pm 0.13$) \citep{Chen}, we have particularly chosen $\beta=0.5$, $0.7$ and $0.9$ for comparison. We can see that the number density profile of the $\beta$-model generally agrees with that predicted by Eq.~(6). The deviation is mainly found at the deep central region $x \le 0.5$ of the hot gas density profile. Note that the observational uncertainties of the hot gas number density at the deep central region are relatively large. Therefore, it is difficult to determine whether Eq.~(6) can better describe the hot gas number density profile at the deep central region than the $\beta$-model. 

Moreover, from Eq.~(11), we have found another universal relation $\rho_sr_s^2=kTC/4 \pi Gm_g$. Since $a=\sqrt{3}$, we get $C \approx 27 \beta/2$ and
\begin{equation}
\rho_s \approx \frac{9\beta kT}{8\pi Gm_gr_c^2}.
\end{equation}
Therefore, we can obtain two analytic expressions, Eqs.~(13) and (14), to relate the NFW profile parameters \{ $r_s$, $\rho_s$ \} (the CDM framework) with the parameters in the $\beta$-model \{ $\beta$, $T$, $r_c$ \} (the hot gas framework) directly. 

We can also consider the effect of the temperature gradient quantitatively. In fact, observations indicate that the average temperature gradients for non-cool-core clusters and cool-core clusters are about $d \ln T/d\ln r \sim 0.08$ and $d \ln T/d\ln r \sim 0.24$ respectively \citep{Hudson}. However, observations also indicate that the temperature profiles outside the central regions slightly decline with radius in galaxy clusters \citep{Vikhlinin,Pratt,Reiprich}. Such temperature decline in the outskirt regions is quite common in galaxy clusters. Strictly speaking, the hot gas temperature profile for each galaxy cluster is somewhat specific. However, some general functional forms can be obtained for further analysis. For example, \citet{Vikhlinin} have obtained a general form of the temperature profile for 13 relaxed galaxy clusters. Later, by analyzing 15 nearby galaxy clusters drawn from a statistically representative sample, \citet{Pratt} have obtained a simple linear model to describe the general hot gas temperature profile for $0.125<r/r_{200}<0.5$:
\begin{equation}
T(r)=T_X\left(1.19-0.74\frac{r}{r_{200}} \right),
\end{equation}
where $T_X$ is the normalized global virial temperature. 

In our analysis, we compare the dark matter profiles between the hot gas framework and the CDM framework at $r=r_s$. Therefore, examining the temperature gradient at $r=r_s$ would be crucial for estimating its potential impact on our study. Following the general form in Eq.~(15) and considering the galaxy clusters with concentration parameter $c=r_{200}/r_s=4-10$, we can get $\gamma=d \ln T(r)/d \ln r=[-0.184,-0.084]$ and $r(d\gamma/dr)=[-0.22,-0.09]$ at $r=r_s$. Putting these values to Eq.~(9) and assuming $a^2 \approx 3$, the percentage change of $\rho_{DM}$ by adding the extra terms of $\gamma$ is 5-12\%. Besides the extra terms, the temperature at $r=r_s$ might also deviate from the normalized global virial temperature $T_X$ for different galaxy clusters. The temperature at $r=r_s$ in Eq.~(15) can be written in terms of $c$: $T(r_s)=T_X(1.19-0.74/c)$. The temperature at $r_s$ for galaxy clusters with $c=4$ and $c=10$ are $T(r_s)=1.005T_X$ and $T(r_s)=1.116T_X$ respectively. Therefore, combining the effects of the extra terms and the temperature deviation at $r=r_s$ for $c=4-10$, the overall percentage increase in $\rho_{DM}$ is 10-13\% if we consider the effect of the temperature gradient. This percentage change is generally smaller than the combined observational uncertainties of $\beta$ and $r_c$ for most of the galaxy clusters (e.g. mean percentage uncertainty of $r_c \approx 13$\% in the sample of \citet{Chen}). Therefore, neglecting the contribution of the temperature gradient in the dark matter density profile is a good approximation. One can also solve Eq.~(1) and Eq.~(9) numerically to obtain a more precise relation between $r_c$ and $r_s$. The resultant value of $a$ would depend on $\beta$ and $T$. Furthermore, the extra term $\gamma$ in Eq.~(4) and the effect of $T(r_s)$ altogether contribute 13-15\% uncertainties for the hot gas density profile. This percentage uncertainty is similar to the one in calculating the dark matter density profile.

Besides, based on the above results, we can derive the cosmological mass-concentration (the relation between the virial mass $M_{\rm vir}$ and the mass concentration parameter $c$) theoretically. Starting from Eq.~(14), we get $\rho_sr_s^2 \propto \rho_sr_c^2 \propto \beta T$. Since observations indicate $M_{\rm vir} \propto T^{1.57 \pm 0.06}$ \citep{Ventimiglia}, we get $M_{\rm vir} \propto (\rho_sr_s^2/\beta)^{1.57 \pm 0.06}$. As mentioned above, the range of $\beta=0.65 \pm 0.13$ is not large so that we simplify the above relation to $M_{\rm vir} \propto (\rho_sr_s^2)^{1.57 \pm 0.06}$. On the other hand, from Eq.~(2), the virial mass at virial radius $r_{200}=cr_s$ can be approximately written as a power-law function of $c$: $M_{\rm vir}=M(r_{200})=4\pi \rho_sr_s^3c^{0.66}$, for the physical range of $c=4-10$ in galaxy clusters. Furthermore, from the definition of the virial mass, we have $M_{\rm vir} \propto r_s^3c^3$. Combining the above power-law relations, we finally get the mass-concentration relation:
\begin{equation}
c \propto M_{\rm vir}^{-0.087^{+0.069}_{-0.074}}.
\end{equation}
Simulations and observations give $c \propto M_{\rm vir}^{-\alpha}$ with $\alpha \approx 0.1$ \citep{Duffy,Dutton,Schaller}. Therefore, our analytic framework can give a good agreement with the simulation and observational results.

\section{Discussion}
In this article, we have applied a simple framework describing the gravitational astrophysics in galaxy clusters. We have derived two important analytic relations that can connect the parameters of cluster hot gas with the parameters of the CDM model: $r_s \approx \sqrt{3}r_c$ and $\rho_s \approx 9\beta kT/8 \pi Gm_gr_c^2$. These two relations imply that the parameters $\beta$ and the core radius of the hot gas density $r_c$ can be uniquely determined by the dark matter scale density $\rho_s$, dark matter scale radius $r_s$ and the average hot gas temperature $T$. In other words, the dark matter distribution and the average hot gas temperature can completely determine the actual hot gas distribution. In view of this, as the hot gas parameters in many galaxy clusters have already been compiled by X-ray studies, we can now quickly obtain the NFW dark matter density profiles for these galaxy clusters by using our two analytic relations.

In fact, some early studies have examined the relations between the CDM model and the hot gas framework. For example, \citet{Suto} have obtained some rough relations connecting the NFW dark matter profile with the properties of the hot gas in galaxy clusters. By assuming a constant temperature profile, they also get the same number density profile shown in Eq.~(6). Later, \citet{Ascasibar,Bode} describe the hot gas profiles of galaxy clusters by using the polytropic gas models and compare the results with that in hydrodynamical simulations. They also get a similar relation shown in Eq.~(10). However, these studies have not derived any explicit analytic relation connecting the observed hot gas parameters with the NFW parameters, such as our most important analytic relation in Eq.~(13). This relation can break the degeneracy of the two parameters $\rho_s$ and $r_s$ appeared in the CDM model, which can allow quick conversion for the hot gas parameters to $\rho_s$ and $r_s$ separately. Moreover, most of these early studies do not have enough information about the hot gas temperature profile. These studies either follow the polytropic gas models \citep{Komatsu,Ascasibar} or fixing the entropy profile \citep{Voit} to constrain the temperature profile, which might involve some extra variables (e.g. the polytropic index) or uncertain functional form of the hot gas temperature. Later, the X-ray data obtained by the {\it Chandra}, {\it XMM-Newton} and {\it Suzaku} are good enough to construct the temperature structure so that we can understand more about the temperature profiles in galaxy clusters \citep{Chen,Reiprich}. We now notice that it is not necessary to involve the polytropic relation or any entropy profile in the hot gas analysis, which can give a more simplified and less uncertain picture. In this study, we revisit the connections between the CDM model and the hot gas framework based on this updated and simplified picture. Surprisingly, we have found two simple and important analytic relations connecting the empirical hot gas parameters and the NFW parameters, which have not been obtained explicitly in previous studies. 

Following this framework, we can also approximately reproduce the hot gas density profile, which have been commonly described by the $\beta$-model in the hot gas framework. As mentioned above, this has been indeed derived in some early studies \citep{Suto}. Since the dark matter density profile (the NFW profile) is universal based on the CDM numerical simulations, this might explain why a universal $\beta$-model could be able to describe the hot gas density profiles in many galaxy clusters (some could be better fitted by the double-$\beta$ model) \citep{Chen}. Besides, we can also use Eq.~(6) to model the hot gas density profile so that we can determine the dark matter parameters more directly.

In deriving the relations, we have a few assumptions involved. First, we assume that hydrostatic equilibrium is achieved in the hot gas. This is a good approximation for relaxed galaxy clusters. A simulation study shows that the systematic error of using hydrostatic mass for galaxy clusters is about 10\% \citep{Biffi}. In particular, the hydrostatic mass bias for cool-core and non-cool-core clusters are 3\% and 13\% respectively \citep{Biffi}. This is due to the fact that clusters are generally not completely relaxed. This systematic effect might not be completely negligible in the view of precision cosmology, especially for describing non-cool-core clusters. More future comparisons between numerical simulations and the hydrostatic model could be useful for removing this systematic effect. Furthermore, the effect of convection may occur in the hot gas medium to affect the hydrostatic equilibrium \citep{Reiprich}. Nevertheless, the density gradients in most of the galaxy clusters are much steeper than the temperature gradients, except a very few cool-core galaxy clusters. Therefore, generally no convection is expected \citep{Reiprich}. Second, we assume that the temperature is almost constant for the hot gas. This assumption is good for many galaxy clusters \citep{Hudson}. However, observations indicate that the temperature profiles of galaxy clusters generally decline with radius outside the central region \citep{Vikhlinin,Pratt,Reiprich}. The temperature can even drop by half when $r$ approaches $r_{200}$. Nevertheless, in our study, we mainly compare the dark matter density profiles of the two frameworks near $r=r_s$ (i.e. $r \sim 0.1-0.3r_{200}$) so that the effect of the temperature decline is less significant. We have also estimated the effect of the temperature decline and it contributes 10-15\% variation compared with the constant temperature scenario considered. This variation is indeed smaller than the observational uncertainties of the hot gas parameters such as $r_c$ and $\beta$. In fact, more precision relations could be obtained by solving Eq.~(1) and Eq.~(9) numerically if necessary. However, no simple analytic relation could be obtained if we have included the temperature gradient terms. Overall speaking, it is still worthwhile to obtain the two simple analytic relations without including the small percentage uncertainty of the temperature gradient terms. 

To conclude, we have derived two analytic relations that connect the hot gas astrophysics with the CDM model, which have not been shown explicitly in previous studies. If one can obtain the dark matter parameters precisely for galaxy clusters by other independent methods, such as gravitational lensing, in addition to precise X-ray observations of the hot gas distribution, then our two analytic relations could be critically examined and verified.

\begin{figure}
\vskip 10mm
 \includegraphics[width=140mm]{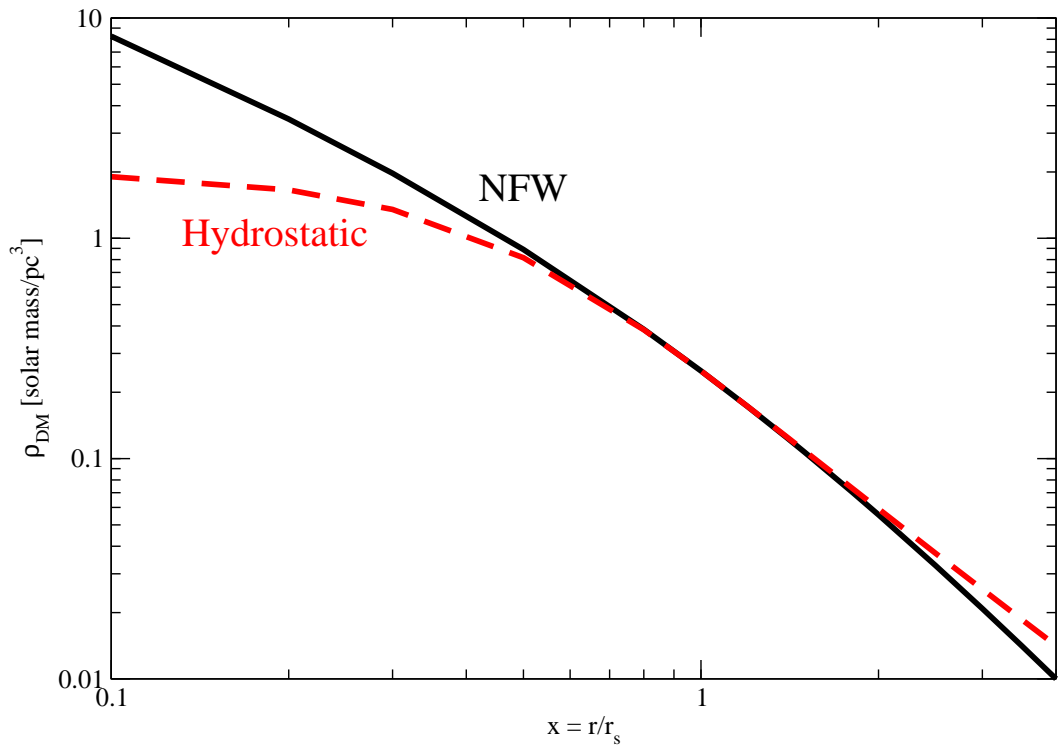}
 \caption{The dark matter density distributions for the NFW profile based on Eq.~(1) (black solid line) and the hydrostatic profile based on Eq.~(9) (red dashed line). Here, we assume $r_s=\sqrt{3}r_c$ and $\rho_s=9 \beta kT/8\pi Gm_gr_c^2=1M_{\odot}$ pc$^{-3}$.}
\vskip 10mm
\end{figure}

\begin{figure}
\vskip 10mm
 \includegraphics[width=140mm]{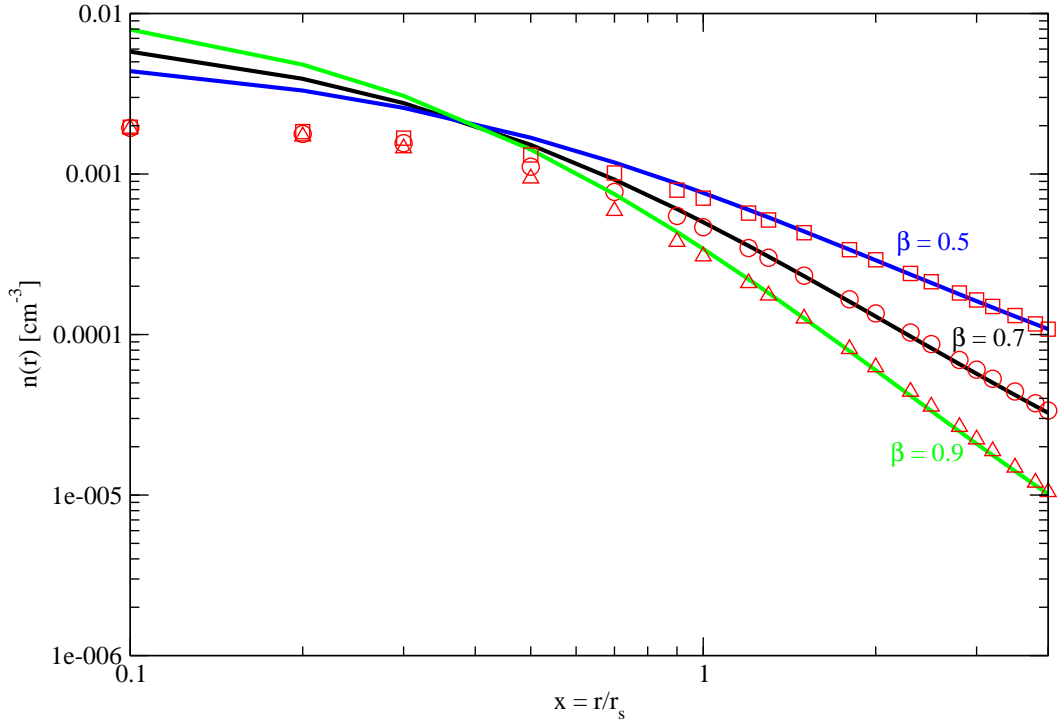}
 \caption{The solid lines (blue, black and green) indicate the number density profiles predicted in our framework. We have used $n_0=0.006$ cm$^{-3}$, $n_0=0.009$ cm$^{-1}$ and $n_0=0.014$ cm$^{-3}$ for $\beta=0.5$, $\beta=0.7$ and $\beta=0.9$ respectively. The squares ($\beta=0.5$), circles ($\beta=0.7$) and triangles ($\beta=0.9$) represent the number density profiles of the $\beta$ model for comparison. A single value of $n_0=0.002$ cm$^{-3}$ is assumed for the $\beta$-model.}
\vskip 10mm
\end{figure}

\section{Acknowledgements}
We thank the anonymous referee for useful constructive feedbacks and comments. The work described in this paper was partially supported by the Seed Funding Grant (RG 68/2020-2021R) and the Dean's Research Fund of the Faculty of Liberal Arts and Social Sciences, The Education University of Hong Kong, Hong Kong Special Administrative Region, China (Project No.: FLASS/DRF 04628).


\begin{thebibliography}{}
\bibitem[Ascasibar et al. (2003)]{Ascasibar} Ascasibar Y., Yepes G., Mueller V. \& Gottloeber S., 2003, Mon. Not. R. Astron. Soc. 346, 731.
\bibitem[Balogh et al. (2011)]{Balogh} Balogh M. L., Mazzotta P., Bower R. G., Eke V., Bourdin H., Lu T. \& Theuns T., 2011, Mon. Not. R. Astron. Soc. 412, 947.
\bibitem[Biffi et al. (2016)]{Biffi} Biffi V. {\it et al.}, 2016, Astrophys. J. 827, 112.
\bibitem[Bode, Ostriker \& Vikhlinin (2009)]{Bode} Bode P., Ostriker J. P. \& Vikhlinin A., 2009, Astrophys. J. 700, 989.
\bibitem[Cavagnolo et al. (2009)]{Cavagnolo} Cavagnolo K. W., Donahue M., Voit G. M. \& Sun M., 2009, Astrophys. J. Supp. 182, 12.
\bibitem[Chan (2014)]{Chan} Chan M. H., 2014, Mon. Not. R. Astron. Soc. 442, L14.
\bibitem[Chan (2020)]{Chan2} Chan M. H., 2020, Phys. Dark Uni. 28, 100478.
\bibitem[Chen et al. (2007)]{Chen} Chen Y., Reiprich T. H., B\"ohringer H., Ikebe Y. \& Zhang Y.-Y., 2007, Astron. Astrophys. 466, 805.
\bibitem[Duffy et al. (2008)]{Duffy} Duffy A. R., Schaye J., Kay S. T. \& Vecchia C. D., 2008, Mon. Not. R. Astron. Soc. 390, L64.
\bibitem[Dutton \& Macci\`o (2014)]{Dutton} Dutton A. A. \& Macci\`o A. V., 2014, Mon. Not. R. Astron. Soc. 441, 3359.
\bibitem[Fong et al. (2018)]{Fong} Fong M., Bowyer R., Whitehead A., Lee B., King L., Applegate D. \& McCarthy I., 2018, Mon. Not. R. Astron. Soc. 478, 5366.
\bibitem[Henden, Puchwein \& Sijacki (2020)]{Henden} Henden N. A., Puchwein E. \& Sijacki D., 2020, Mon. Not. R. Astron. Soc. 498, 2114.
\bibitem[Hudson et al. (2010)]{Hudson} Hudson D. S., Mittal R., Reiprich T. H., Nulsen P. E. J., Andernach H. \& Sarazin C. L., 2010, Astron. Astrophys. 513, A37.
\bibitem[Jauzac, Harvey \& Massey (2018)]{Jauzac} Jauzac M., Harvey D. \& Massey R., 2018, Mon. Not. R. Astron. Soc. 477, 4046.
\bibitem[Komatsu \& Selijak (2001)]{Komatsu} Komatsu, E. \& Seljak U., 2001, Mon. Not. R. Astron. Soc. 327, 1353.
\bibitem[Meneghetti et al. (2020)]{Meneghetti} Meneghetti M. {\it al.}, 2020, Science 369, 1347.
\bibitem[Munari et al. (2016)]{Munari} Munari E. {\it et al.}, 2016, Astrophys. J. 827, L5.
\bibitem[Navarro, Frenk \& White (1997)]{Navarro} Navarro J. F., Frenk C. S. \& White, S. D. M., 1997, Astrophys. J. 490, 493.
\bibitem[Pratt et al. (2007)]{Pratt} Pratt G. W., B\"ohringer H., Croston J. H., Arnaud M., Borgani S., Finoguenov A. \& Temple R. F., 2007, Astron. Astrophys. 461, 71.
\bibitem[Qiu et al. (2020)]{Qiu} Qiu Y., Bogdanovi\`c T., Li Y., McDonald M. \& McNamara B. R., 2020, Nat. Astron. 4, 900.
\bibitem[Reiprich \& B\"ohringer (2002)]{Reiprich2} Reiprich T. H. \& B\"ohringer H., 2002, Astrophys. J. 567, 716.
\bibitem[Reiprich et al. (2013)]{Reiprich} Reiprich T. H., Basu K., Ettori S., Israel H., Lovisari L., Molendi S., Pointecouteau E. \& Roncarelli M., 2013, Sp. Sci. Rev. 177, 195.
\bibitem[Schaller et al. (2015)]{Schaller} Schaller M. {\it et al.}, 2015, Mon. Not. R. Astron. Soc. 451, 1247.
\bibitem[Suto, Sasaki \& Makino (1998)]{Suto} Suto Y., Sasaki S. \& Makino N., 1998, Astrophys. J. 509, 544.
\bibitem[Ventimiglia, Voit \& Rasia (2012)]{Ventimiglia} Ventimiglia D. A., Voit G. M. \& Rasia E., 2012, Astrophys. J. 747, 123.
\bibitem[Vikhlinin et al. (2006)]{Vikhlinin} Vikhlinin A., Kravtsov A., Forman W., Jones C., Markevitch M., Murray S. S. \& Van Speybroeck L., 2006, Astrophys. J. 640, 691.
\bibitem[Voit et al. (2002)]{Voit} Voit G. M., Bryan G. L., Balogh M. L. \& Bower R. G., 2002, Astrophys. J. 576, 601.
\bibitem[Zhang et al. (2006)]{Zhang1} Zhang Y.-Y., B\"ohringer H., Finoguenov A., Ikebe Y., Matsushita K., Schuecker P., Guzzo L. \& Collins C. A., 2006, Astron. Astrophys. 456, 55.
\bibitem[Zhang et al. (2007)]{Zhang2} Zhang Y.-Y., Finoguenov A., B\"ohringer H., Kneib J.-P., Smith G. P., Czoske O. \& Soucail G., 2007, Astron. Astrophys. 467, 437.
\end{thebibliography}
\end{document}